\documentclass[prl,twocolumn,aps,showpacs,amsmath,amssymb]{revtex4}

\usepackage{graphicx}
\usepackage{dcolumn}
\usepackage{amsmath}

\begin{document}

\title{Ultrafast Enhancement of Ferromagnetism via Photoexcited Holes in GaMnAs}
\author{J. Wang,$^{1\dagger}$ I. Cotoros,$^{1}$ K. M. Dani,$^{1}$
X. Liu,$^2$ J. K. Furdyna,$^2$ and D. S. Chemla$^{1}$}

\affiliation{$^1$Materials Sciences Division,
E.O. Lawrence Berkeley National Laboratory and Department of Physics, University of California
at Berkeley, Berkeley, California
94720, U.S.A.\\
$^2$Department of Physics, University of Notre Dame, Notre Dame, Indiana 46556, U.S.A.}

\date{\today}

\begin{abstract}

We report on the observation of {\em ultrafast photo-enhanced ferromagnetism}
in GaMnAs. It is manifested as a transient magnetization increase on a 100-ps time scale, after an
initial sub-ps demagnetization.
The dynamic magnetization enhancement exhibits a maximum below the Curie temperature T$_c$ and dominates
the demagnetization component when approaching T$_c$.
We attribute the observed ultrafast collective ordering to
the $p$-$d$ exchange interaction between
photoexcited holes and Mn spins, leading to a correlation-induced peak around 20K and a transient increase in T$_c$.

\end{abstract}
\maketitle

There has been long and intense interest in searching for possibilities for {\em ultrafast enhancement}
of collective magnetic order via photoexcitation. Such photoexcitation would lead to fascinating
opportunities both for establishing a transient cooperative phase from an uncorrelated ground
state and for determining the relevant time scales for the build-up of order parameters.
Previous time-resolved investigations in paramagnetic II-VI semiconductors provided evidence
for collective photoexcitations that led to the formation of magnetic
polarons \cite{HarrisetalPRL1983}. Substantial recent
progress has been made in observing ultrafast spin reorientations, e.g., in
antiferromagnetic materials \cite{KimeletAl04Nature, KimeletAl05Nature}.
However, most prior experiments in magnetically-ordered materials only show an
{\em ultrafast decrease} of the magnetization amplitude due to laser induced
electronic heating \cite{BeaurepaireetalPRL96}. Recently, several
experiments in strongly correlated manganites and transition metal
alloys revealed transient photo-induced magnetization on
ultrafast time scales, but only {\em demagnetization} could be seen at temperatures away from
T$_c$. The role of ultrafast pumping is most likely a {\em thermal} perturbation of
competing phases near critical points \cite{McGilletalPRB2005, JuetalPRL2004}.

The discovery of hole-mediated ferromagnetism in III-V ferromagnetic semiconductors (III-V FMSs) such
as GaMnAs and InMnAs \cite{Munekata} offers unique opportunities and
flexibility for {\em nonthermal} control of magnetism.  Unlike other types of
magnets, the ferromagnetic exchange between localized
Mn moments is mediated by free hole spins through $p$-$d$ interaction
$H_{p-d}\sim J_{pd}\mathbf S\cdot \mathbf s$ ($J_{pd}$ $\sim$ 1 eV),
making the magnetic properties a sensitive function of the hole density.
For instance, the trend of experimental T$_c$ in GaMnAs
is shown
to be proportional to $p^{1/3}J_{pd}^{2}$, for a wide
range of hole densities $p$ \cite{MacDonaldetal2005}. Recently, hole-density-tuning
via external stimuli such as CW light excitation \cite{Koshiharaetal97PRL} and
electrical gating \cite{ohnoetal2000} has demonstrated clear enhancement of
magnetization and increase of T$_c$, as illustrated in Fig. 1(a).
However, no time-resolved experiments
in III-V FMSs have shown transient photo-induced magnetization, and the time scale for the enhancement of
collective order is
completely unknown.
So far only ultrafast demagnetization and quenching dynamics were observed,
due in part to the relatively high pump fluences used ($\sim$ 1 mJ/cm$^2$) \cite{KojimaetAl03PRB,wangetalPRL2005}.
Moreover, recent theory has pointed to the critical role of the Mn-hole exchange
correlation in {\em ultrafast, nonthermal}
manipulation of magnetization in GaMnAs \cite{ChovanetalPRL2006}, but no experimental
evidence for this has been reported.
In addition to their fundamental scientific interest,
III-V FMSs are among the most promising
candidates for future "multifunctional" devices and for quantum information technology based on the
spin degree of freedom \cite{Awschalometalbook}. It is thus of significance to explore ferromagnetic
enhancement in III-V FMSs to technologically
important, sub-ns time scales.

In this Letter, we report the observation of {\em ultrafast enhancement
of ferromagnetism} in GaMnAs.  Our data clearly show photo-induced magnetization on a 100-ps time scale
after initial sub-ps demagnetization.
The dynamic magnetization enhancement exhibits a maximum below T$_c$ and
dominates the demagnetization component when approaching T$_c$.
Our analysis and theoretical simulations based
on the $H_{p-d}$ interaction between photoexcited holes and Mn spins
explain the salient features of the experiment showing, in particular, a correlation-induced
peak around 20K and a transient increase in T$_c$.

The sample studied in the present work was a GaMnAs/GaAs heterostructure with a Curie temperature of 77 K.
 The sample was grown by low-temperature molecular beam epitaxy (MBE) and consisted of a
 73-nm Ga$_{0.925}$Mn$_{0.075}$As layer deposited on a GaAs buffer layer and a
 semi-insulating GaAs (100) substrate.
The background hole density was approximately $3 \cdot 10^{20}$ cm$^{-3}$.
Our measurements involved ultraviolet (UV) pump/near-infrared (NIR) magneto-optical Kerr effect (MOKE) spectroscopy.
The experimental setup consisted of a femtosecond oscillator with 120 fs pulse duration and
 a BBO crystal in the pump path for doubling the photon energy.
 The pump beam was at 3.1 eV and was linearly polarized, with peak fluences 
$\simeq$ 10$\mu$J/cm$^2$. 
Fig. 1(b)
 illustrates that fs pump pulses create
 a large density of holes in the valence band of GaMnAs.
 A small fraction of the fundamental beam at 1.55 eV was used as a probe, detecting magnetization
 via the polar MOKE angle $\theta_K$\cite{wangetalreview2006}.
The low pump peak fluences and the high
pump photon energy minimize spurious
 effects such as two-photon
 absorption and pump
scattering.  Additionally, the "magnetic origin" of the transient MOKE response is confirmed
by separate measurements showing the overlap of the pump-induced rotation and ellipticity through
the entire time scan range \cite{KoopmansetAl00PRL}.

\begin{figure}[floatfix]
\includegraphics [scale=0.35] {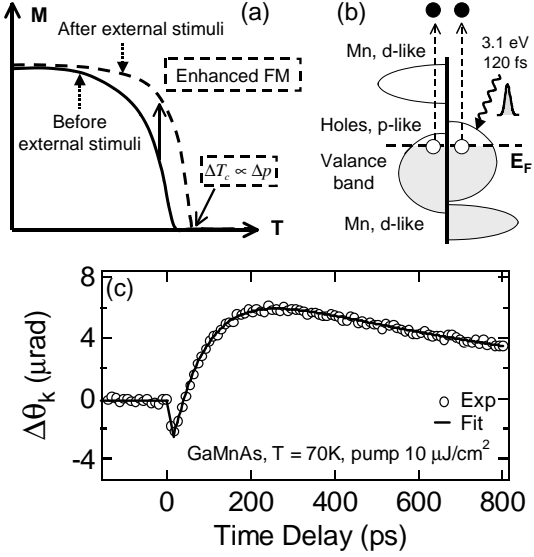}  
\caption{(a) Illustration of hole-density-tuning effects via
external stimuli in III-V FMSs seen in the
static experiments \cite{ohnoetal2000,Koshiharaetal97PRL}.
FM: ferromagnetism. $\triangle p$ is hole density change.
(b) Schematic diagram of the spin-dependent density of states in GaMnAs.
 fs pump pulses create a transient population of
 holes in the valence band.
 (c) Time-resolved MOKE dynamics at 70K and under
 1.0T field. Transient enhancement of magnetization, with
 $\sim$ 100 ps rise time, is clearly seen after initial
 fast demagnetization.
Thick line is the fit described in the text.}
\label{typical}
\end{figure}

A typical temporal profile of transient MOKE changes $\Delta\theta_K$ at 70K is shown in Fig. 1(c),
with a field of 1.0T perpendicular to the sample surface to align the magnetization. Two
mutually competing dynamic magnetization processes are observed: an initial sub-ps
demagnetization ($\Delta\theta_K<0$), followed by a distinct magnetization
rise on a 100 ps time scale ($\Delta\theta_K>0$). 
The two processes clearly show different temperature dependences, as shown in Fig. 2(a)-(b).  At elevated
lattice temperature, the 200 fs demagnetization components [inset of Fig. 2(b)] quickly diminish and nearly disappear above T$_c$.
This is also seen in the 600 fs and 3 ps traces in Fig. 2(c).
More intriguingly, an enhancement of the transient magnetization begins
to dominate the demagnetization component at high temperatures.
For instance,
the net magnetization changes become positive above 40K at long time delays, e.g., at 240 ps [Fig. 2(c)].
The photoinduced magnetization persists above T$_c$ - as is clearly visible in the
80K trace [Fig. 2(b)] - and gradually vanishes at higher temperatures.

\begin{figure}
\includegraphics [scale=0.5] {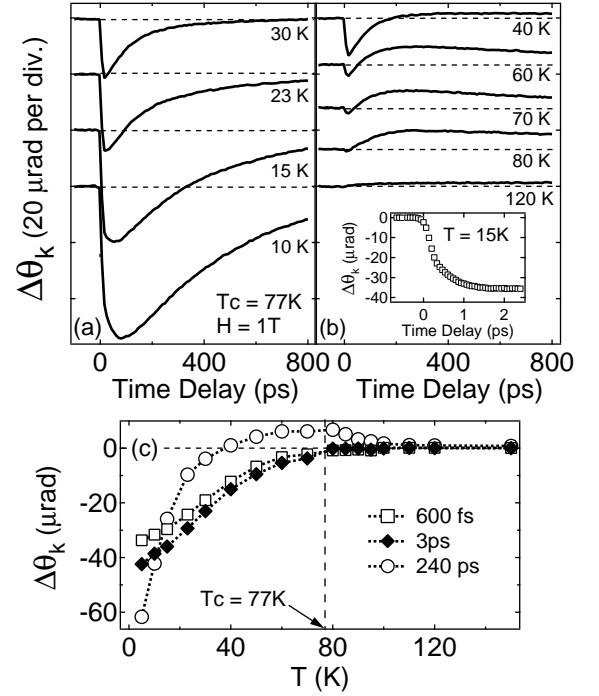}  
\caption{(a)-(b) Temporal traces of photoinduced magnetization changes
at different temperatures. All traces are intentionally offset for clarity.
Inset: the first 2 ps demagnetization dynamics at 15 K.
(c) Temperature dependence of magnetization changes at different time delays - 600 fs (squares),
3 ps (diamonds) and 240 ps (circles), respectively.}
\label{mag-dep}\end{figure}

MOKE signals, measured with the 1.55 eV probe, arise from the macroscopic magnetization $\mathbf{M}$ -
average localized Mn spins - through the coupling of the spin-split electronic states near the band edge.
 The background carrier spin contribution to $\mathbf{M}$ is negligible, and
 photoexcited transient carriers are not spin polarized since the pump contains no net angular momentum.
 Thus the positive MOKE signals, rising on a 100-ps time scale, clearly
 indicate an ultrafast alignment of Mn spins
and an enhancement of ferromagnetic order.  Our results reveal
the ultrafast time scale of this process, not accessible in
previous static measurements \cite{Koshiharaetal97PRL,ohnoetal2000}.

We attribute the observed {\em ultrafast
photo-enhanced ferromagnetism} to the transient hole-Mn
interaction via the $H_{p-d}$ exchange, as follows. At early
pump-probe delays ($\triangle t$ $\sim$ 0 fs), the ultrashort
laser pulses generate a nonequilibrium distribution of
spin-unpolarized
electron-hole pairs in GaMnAs under a finite external H field.
 During the first
ps ($\triangle t<1ps$), the photoexcited hot holes will experience efficient spin-flip scattering
with the localized Mn moments, manifesting this as a sub-ps demagnetization component.
This results from the off-diagonal elements of the exchange
Hamiltonian $H_{p-d}$ ($\sim$$J_{pd}\mathbf S_{\pm }\cdot \mathbf s_{\mp }$), which
cause the spin polarization of the Mn ions to transfer to the holes within several 100s of fs,
similar to the fs demagnetization first reported in InMnAs \cite{wangetalPRL2005}.
Meanwhile, the hot hole distribution quickly
cools via carrier-phonon scattering (optical phonon energy $\sim$36 meV), resulting in a rapid termination
of demagnetization (within the first ps).
At longer pump-probe delays of $\triangle t>1ps$, the photoexcited, thermalized holes,
settling down in the spin-split bands,
can now participate in the process of
hole-mediated ferromagnetic ordering.  These {\em extra} holes
enhance the Mn-Mn exchange correlation and polarize Mn spins via
the mean-field (diagonal) elements of the exchange Hamiltonian $H_{p-d}$ ($\sim$$J_{pd}\mathbf S_{z}\cdot \mathbf s_{z}$),
thereby increasing the macroscopic
magnetization.

\begin{figure}
\includegraphics [scale=0.65] {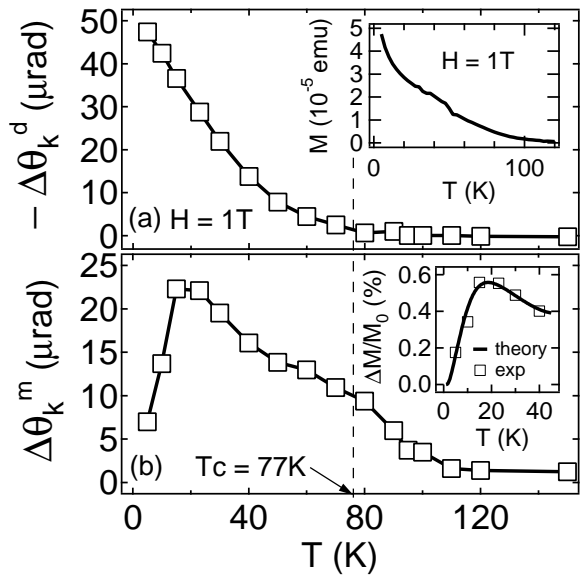}  
\caption{Decomposed  demagnetization $-\triangle \theta _{k}^{d}$ [(a)] and
enhanced magnetization $\triangle \theta _{k}^{m}$ [(b)] components are
 plotted as a function of temperature.
The static magnetization curve under 1.0T field is shown as inset in (a). The simulation
$\triangle M/M_{0}$ (thick line, $\triangle T_{c}= 1.1K $ and the
 hole-Mn ratio of 0.06) and experimental
values (circles, normalized by static $\theta _{k}$ at 5K) of the
photo-enhancement peak $\sim$ 20K are shown as inset in (b).}
\label{power}\end{figure}

In order to elucidate the salient features of the {\em photo-enhanced ferromagnetism},
we decompose the transient MOKE changes shown in Fig. 2 into demagnetization ($-\triangle \theta _{k}^{d}$)
and enhanced magnetization ($\triangle \theta _{k}^{m}$) components, based on
their different time scales. The temporal profile of $\triangle \theta _{k}$ is well described by
$\triangle \theta _{k}^{d}\cdot exp(-t/\tau _{d})+\triangle \theta _{k}^{m}\cdot
(1-exp(-t/\tau _{m}))\cdot exp(-t/\tau _{c})$.
Here $-\triangle \theta _{k}^{d}$ and $\tau _{d}$ in the first term are the magnitude of demagnetization and the
recovery time determined by the slow heat diffusion process, respectively.
In the second term, $\triangle \theta _{k}^{m}$ and $\tau _{m}$ are the magnitude
and build-up time of the enhanced
magnetization component, respectively, while $\tau _{c}$ accounts for the final decay of the magnetization enhancement via
hole diffusion and recombination.
The time constant $\tau _{c}$ is on the
order of a few ns, as seen in the decay of the positive MOKE signal.
As an example, the thick line in Fig. 1(c) represents the fit of the MOKE dynamics at 70K.

Fig. 3 plots the temperature dependence
of $-\triangle \theta _{k}^{d}$ and $\triangle \theta _{k}^{m}$.
The $-\triangle \theta_ {k}^{d}$ profile resembles the static
magnetization curve of the sample, exhibiting strong
deviation from the classical mean-field convex shape [inset, Fig. 3(a)].
This non-classical behavior of magnetization arises from the
existence of two strongly interacting
spin ensembles, Mn and holes, as discussed in \cite{sarmaetalPRB2003}.
More intriguingly, the magnetization enhancement $\triangle \theta _{k}^{m}$ in Fig. 3(b)
shows a distinctly different temperature profile, with a peak of $\sim 0.5$\% of the
saturation magnetization $M_{0}$ around 20K (static $\theta _{k}$ at 5K $\sim$ 4 mrad) and a prolonged
tail extended to $\sim$ 120K.
The extended profile beyond T$_c$ is expected as the combined effect
of an applied external field of 1.0T and of hole-enhanced magnetic susceptibility in the paramagnetic state.
The most salient feature of the {\em photo-enhanced
ferromagnetism} is the peak near 20K, which is a manifestation of the Mn-hole correlation $H_{p-d}$.
This can be qualitatively understood as follows: the ferromagnetic molecular field acting on the Mn ions (holes) is
determined by the average spin polarizations of the holes (Mn ions) via the
Mn-hole exchange coupling J$_{pd}$. The effective field
acting on the holes is much larger than that acting on the Mn
 ions, because of the large density of Mn ions compared to the holes.
As a consequence, the hole polarization will remain
saturated at a temperature $T_h$ much higher than that needed for
saturating the Mn magnetization ($T_{Mn}$$\sim$0).
As the lattice temperature increases above $T_{Mn}$ but
less than $T_h$, the Mn spins with partial alignment begin to be efficiently
polarized via photoexcited holes with near-unity magnetization.
However, as the lattice temperature
rises higher than $T_h$, because of reduced hole polarization the efficiency of this magnetization
enhancement process quickly
drops, thus resulting in a magnetization enhancement maximum at
some temperature on the order of $T_h$.

Next we present a simple theoretical calculation to simulate the Mn-hole correlation-induced peak around 20K.
As we discussed, the hole-enhanced ferromagnetic correlation results in an increase of T$_c$ ($\triangle T _{c}>0$).
We calculate the
 non-classical magnetization curve [inset, Fig. 3(a)] based on a modified Weiss
 mean-field model to take into account the $H_{p-d}$ correlation \cite{sarmaetalPRB2003}:
\begin{eqnarray*}
M(T,\Delta T_{c})/M_{0}=B_{S}[-3{T_{c}+\Delta T_{c}\over T}
\gamma S^{\ast}\times B_{s}[
-3{T_{c}+\Delta T_{c}\over T}
\\ \gamma ^{-1}S^{\ast}{M(T,\Delta T_{c})\over M_{0}}
+{g_{h}\mu _{B}H\over k_{B}T}
]-{g_{i}\mu _{B}H\over k_{B}T}]\end{eqnarray*}where
B$_{S,s}$(x) is the Brillouin function, $\gamma $ is the square
root of the hole and Mn density ratio $\sqrt {p/n_{i}}$,
g$_i$(g$_h$) is Mn(hole) g factor and S$^{\ast}$ is given by $\sqrt {sS/(s+1)(S+1)}$.
The Mn and hole spins, S and s, are 5/2 and 1/2, respectively.
At each given $\triangle T _{c}$, the magnetization enhancement $\triangle M/M_{0}$ is solved
self-consistently as a function of lattice temperature.  The calculated temperature
dependent $\triangle M/M_{0}$ and experimental $\triangle \theta _{k}^{m}$ (normalized by static $\theta _{k}$ at 5K)
up to 40K are shown in the inset
in Fig. 3 (b) for $\triangle T_{c}=1.1K$.
 The results of the calculation compare
well with the experimental photo-enhancement peak.
In addition, we can also estimate $\triangle T_{c}$ using an analytical expression
derived from the Zener Model, $\triangle T_{c}={1\over 3}
T_{c}\times \triangle p/p$ \cite{dietletalSci2000}.
Knowing the ratio of photoexcited and background hole
densities $\sim 2\%$, we estimate $\triangle T_{c}\sim$ 0.5K, in
reasonable agreement with the value used in the simulation.

\begin{figure}
\includegraphics [scale=0.65] {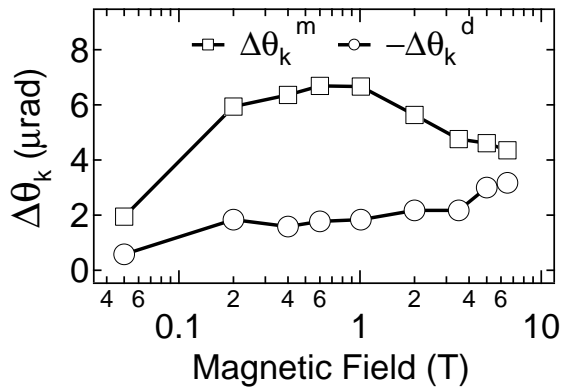}  
\caption{
Magnetic field dependence of photo-induced magnetization
enhancement $\triangle \theta _{k}^{m}$ and
 demagnetization $\triangle \theta _{k}^{d}$ at 78K.}
\label{temperature}\end{figure}

The photoexcited transient electrons have little effect on
$\Delta\theta_K$, because of the short free electron lifetime
($<$ 1 ps) \cite{wangetalPRB2005}. In addition, the standard
thermal demagnetization via heat transfer from phonons,
similar to the slow demagnetization component reported in \cite{KojimaetAl03PRB,wangetalPRL2005},
is also seen at temperatures below 15K [Fig. 2(a)], which is not important
to the physical picture discussed in this paper.

So far our discussion has been based on a situation involving a
fixed magnetic field of 1.0T, which is important at low temperatures in order to eliminate other
distractions (such as magnetization reorientation effects),
without losing universality.  It is worthwhile, however, to consider the
field dependence of the magnetization enhancement at temperatures just above T$_c$ and possible {\em transient} signatures
of photo-induced critical phenomena, e.g., para- to ferromagnetic phase
transition via the pump-induced increase of T$_c$.
Fig. 4 presents the field dependence of
the magnetization enhancement $\triangle \theta _{k}^{m}$ and
demagnetization $-\triangle \theta_{k}^{d}$ at 78 K.
Here the most interesting aspects lie in the fact that $\triangle \theta_{k}^{m}$
 is relatively constant across a wide field range (0.2T - 1T)
and only drops off close to zero T. It is crucial to note that, unlike static
equilibrium measurements, {\em transient} pump-induced magnetization above $T_{c}$ always
reduces to zero at small external field, even though $\triangle T_{c} >0$.  A below-threshold H field
is not able to activate subsequent growth of magnetic
domains, even though they are nucleated via possible photo-induced long-range
ferromagnetic correlation.
Therefore, although the 1K increase of $\triangle T_{c}$ and the substantial
transient magnetization enhancement at a small field of 0.05T suggest a transient photo-induced para- to ferromagnetic
phase transition, more studies are needed to further elucidate the details of these transient
features.  Finally, the decrease (increase) of $\triangle \theta _{k}^{m}$($-\triangle \theta _{k}^{d}$) observed
from 2T to 7T is expected from the larger static magnetization achieved at higher
fields, resulting in the smaller photo-induced changes, as indeed shown by the data.

In summary, we have observed {\em ultrafast photo-enhanced ferromagnetism}
in GaMnAs.  Our data clearly show that the dynamic magnetization build-up occurs
on a 100-ps time scale and exhibits a maximum below $T_c$.
Our analysis and theoretical simulations, based on $H_{p-d}$ interaction between photoexcited holes and Mn spins,
explain the salient features of the experimental observations, demonstrating in particular a correlation-induced
peak below T$_c$ and a
transient increase of T$_c$.
Our
measurements thus reveal a new transient collective magnetic phenomenon,
and identify the critical role of {\em non-thermal} Mn-hole exchange correlation
in this photo-induced cooperative behavior.
The new functionalities in sub-ns time scales reported here may open
future opportunities for high-speed spin-photon-charge integrated devices.

\smallskip

This work was supported by the Office of Basic Energy Sciences of the US Department of
Energy and by the National Science Foundation.  We thank B. A. Schmid, R. A. Kaindl and {\L}. Cywi{\'n}ski
for illuminating discussions.

\medskip

\noindent$^{\dagger}$To whom correspondence should be addressed.
Electronic address: JWang5@lbl.gov.



\end{document}